\documentclass[pra,aps,twocolumn,floatfix,showpacs]{revtex4-1}

\usepackage{graphicx} 

\usepackage{amsmath,amssymb}

\begin{document}

\title{Scaling laws for the non-linear coupling constant of a
Bose-Einstein condensate at the threshold of delocalization}

\author{R. Cabrera-Trujillo}
\email{trujillo@fis.unam.mx}
\affiliation{Instituto de Ciencias F\'isicas, Universidad Nacional 
Aut\'onoma de M\'exico, Ap. Postal 48-3, Cuernavaca, Morelos, 62251, M\'exico}

\author{M. W. J. Bromley}
\affiliation{Centre for Quantum Atom Optics, School of Mathematics and Physics,
The University of Queensland, Brisbane, Queensland, 4072, Australia}

\author{B. D. Esry}
\affiliation{Department of Physics, Kansas State University, Manhattan,
Kansas 66506, USA}

\pacs{03.75.-b, 
      03.75.Hh, 
      05.30.Jp, 
      67.85.-d, 
      67.85.Bc  
}

\begin{abstract}
We explore the localization of a quasi-one-, quasi-two-, and three-dimensional
ultra-cold gas by a finite-range defect along the corresponding 
'free'-direction/s.
The time-independent non-linear Schr{\"o}dinger equation that describes a
Bose-Einstein condensate was used to calculate the maximum non-linear coupling
constant, $g_{max}$, and thus the maximum number of atoms, $N_{max}$,
that the defect potential can localize.  
An analytical model, based on the Thomas-Fermi approximation, is 
introduced for the wavefunction.   
We show that $g_{max}$ becomes a function of $R_0\sqrt{V_0}$ for various 
one-, two-, and three-dimensional defect shapes with depths $V_{0}$ and 
characteristic lengths $R_{0}$.  
Our explicit calculations show surprising agreement with this crude model 
over a wide range of $V_{0}$ and $R_{0}$.  
A scaling rule is also found for the wavefunction for the ground state at 
the threshold at which the localized states approach delocalization.
The implication is that two defects with the same product $R_{0}\sqrt{V_0}$ 
will thus be related to each other with the same $g_{max}$ and will have the 
same (reduced) density profile in the free-direction/s.
\end{abstract}

\maketitle 

\section{Introduction}

Since the experimental realization of Bose-Einstein condensates (BEC) in 1995 
\cite{Anderson-MH95-269science198,Davis-KB95-75prl3969}, interest in the 
physics of ultracold atoms has grown and new areas of research have emerged.
At the same time, it has renewed the interest in studying the collective
dynamics of macroscopic ensembles of atoms occupying the same single-particle
quantum state \cite{Griffin-A95-BEC,Parkins-AS98-303pr1,Dalfovo-F99-71rmp463}. 
This, in turn, has created the need for new technology to study ultracold atoms. 
There are two technologies that motivate our present study;
one is the {\em atom chip}, the second is {\em sculptured} optical
field-based atom trapping.  Both of which are capable of generating a
myriad of trap geometries.

Our fundamental goal in this work is to determine the general scaling law that 
determines the maximum number of atoms that can be trapped by
attractive defects described by 1-D, 2-D, and 3-D potentials
when combined with confining harmonic atom traps.  
This corresponds to determining the limit at which the system transitions 
from a bound to a scattering state.

Our first motivation are the atom chips which play an important role in 
atomic physics, enabling the cooling and trapping of a BEC in a waveguide 
which is created by magnetic fields generated above patterned micro-wire 
circuits \cite{Fortagh-J07-79rmp235}.
Atom chips have already enabled the study of matter-wave interference
phenomena \cite{Cronin-AD09-81rmp1051}, and may improve other atomic 
measurement devices such as atomic clocks in the future.
Ideally, the BEC is loaded/trapped in the transverse ground state,
but is allowed free-space propagation along the third dimension.  
An ultra-cold wave-packet can then be transported through quasi-1-D waveguides 
due to the strong confinement in the two transverse dimensions 
\cite{Leanhardt-AE02-89prl040401}.

Our second motivation is the experimental realization of potential traps
with different shapes that have been achieved using a rapidly moving laser
beam that {\em paints} a time-averaged optical dipole potential.
There, a BEC is created in arbitrary geometries
\cite{Schnelle-SK08-16oe1405,Henderson-K09-11njp043030}.
Effectively, the BEC confinement can be strong in one-dimension, reducing the
BEC to be quasi-2-D in shape.  
The BEC can then be manipulated in these two-dimensions using a time-averaged 
potential.
Both the atom chip and sculptured traps motivate us to explore a variety
of shapes for a basic trap which includes a defect potential.

In the study of the structure of a gas of ultra-cold atoms, it is necessary to
account for the interaction of the $N$ atoms.  
That is, each atom moves in an average field due to the other $N-1$ atoms 
that surrounds it. 
The properties of a BEC at $T=0$ is well-described by a mean-field
approximation which results in a non-linear Schr{\"o}dinger equation (NLSE)
for the single-particle orbitals \cite{Esry-BD97-55pra1147}
\begin{equation}
  \label{eqn:gpe}
  i\frac{\partial\:}{\partial t}\Psi(\mathbf{r},t)
  = \left[-\frac{1}{2} \nabla^2 +V(\mathbf{r})+
g_3\lvert\Psi(\mathbf{r},t)\rvert^2\right]\Psi(\mathbf{r},t).
\end{equation}
The nonlinear coupling constant characterizes the short-range pairwise 
interactions, and is given for $N$ bosons by $g_3=4\pi {a_{s}}(N-1)/a_{\perp}$
according to (number conserving) Hartree-Fock theory \cite{Esry-BD97-55pra1147}.
This constant depends on the {\em s}-wave atom-atom scattering length, 
$a_{s}$, of two interacting bosons.  
Note that alternative treatments give $g_3 \propto N$, and then the above 
NLSE is known as the Gross-Pitaevskii equation (GPE) 
\cite{Dalfovo-F99-71rmp463}.
Eq.~(\ref{eqn:gpe}) has been written in oscillator units (o.u.), with lengths 
$a_{\perp} =\sqrt{\hbar/m\omega_{\perp}}$ for a chosen frequency 
$\omega_{\perp}$ and $m$ being the mass of the individual atoms that 
compose the BEC.  
The energy is thus given in units of the oscillator energy 
$\hbar\omega_{\perp}$, and time in units of $1/\omega_{\perp}$.
Back in S.I. units, $g_3^{SI} = 4\pi \hbar^2 a_s (N-1) / m = \hbar^2 
g_3/(a_\perp m)$.

For the present study, we chose time-independent potentials for 
$V(\mathbf{r})$ with one of three structures:
\begin{eqnarray}
  \label{eqn:pots} 
    V_1(x,y,z) &=& \frac12 m \omega_\perp(y^2 + z^2) + V^\prime_1(x)\:, \\
    V_2(x,y,z) &=& \frac12 m \omega_\perp(z^2) + V^\prime_2(x,y)\:, \\
    V_3(x,y,z) &=& V^\prime_3(x,y,z) \:.
\end{eqnarray}
That is, we partition the defect potential $V^\prime$, which is
of interest here, from the transverse harmonic trap potential applied
at a frequency $\omega_{\perp}$.

The $V^\prime_1$ defect potentials, for example, can be generated by a 
local modification of the transverse waveguide confinement, such as a 
constriction
\cite{Jaaskelainen-M02-66pra023608,Lahaye-T03-8cnsns315,Leboeuf-P03-68pra063608,Koehler-M05-72pra023603}
or a local curvature
\cite{Leboeuf-P01-64pra033602,Bromley-MWJ03-68pra043609,Bromley-MWJ04-69pra053620}. 
Whether a defect potential acts as an obstacle or a sink, a wave-packet will 
interfere with, and possibly lose atoms as it goes through the defect
due to the non-linearity \cite{Ernst-T10-81pra033614,Gattobigio-GL10-12njp085013},
changing the interaction for any subsequent atom.
In general, propagation through a perturbation in a 1-D waveguide results in 
unwanted transverse excitations of the BEC 
\cite{Leanhardt-AE02-89prl040401,Bromley-MWJ04-70pra013605}.

The presence of a transverse-$\omega_\perp$ potential allows for the reduction
of the NLSE from 3-D to either a 2-D or 1-D form.  
In the 1-D, or $V_1$, case we assume that the tight-waveguide limit applies 
where the longitudinal-($x$) size of the BEC is larger than its 
transverse-($y,z$) cross section.  
This allows for us to integrate out the transverse dimensions which are 
energetically frozen in the harmonic ground state. 
This results in a 1-D NLSE,
\begin{equation}
  \label{eqn:1dgpe} 
  i\frac{\partial\:}{\partial t}\Psi(x,t)
    = \left[-\frac{1}{2} \frac{\partial^2\:}{\partial x^2}+V^\prime_1(x)+g_1\vert\Psi\vert^2\right]\Psi(x,t) 
\end{equation}
with an effective 1-D coupling constant, $g_1 = g_3/(2\pi) = 2 
{a_{s}}(N-1)/a_{\perp}$.
Back in S.I. units $g_1^{SI}=g_3^{SI}/(2\pi a_\perp^2)$, and note that 
the intermediate reduction from 3-D to 2-D gives a similar NLSE with constant 
$g_2 = g_3/\sqrt{2\pi}$.

By systematically adding more and more atoms with $a_{s} > 0$ into the 
potential, at some $g_{3,max}$ the total 3-D eigenenergy, $\varepsilon_3$, 
of the system becomes null with respect to the transverse trap energy 
\cite{Leboeuf-P01-64pra033602}.
For the three geometries in Eq.~(\ref{eqn:pots}), this corresponds
to $\varepsilon_3 = \hbar \omega_\perp$,
$\varepsilon_3 = \frac12 \hbar \omega_\perp$ or $\varepsilon_3 = 0$.
Determining the $g_{3,max}$ is non-trivial since at this point the system
is no longer bound, the delocalized atoms can reach $|\mathbf{r}| = \infty$,
and thus cannot be represented by a square-integrable wavefunction,
and needs to be considered as the limit $\varepsilon_3 \to 0$ from below.

The first analytical approximations for $g_{1,max}$ were simultaneously
derived by Carr {\em et al.} \cite{Carr-LD01-64pra033603} and
Leboeuf and Pavloff \cite{Leboeuf-P01-64pra033602}.
Carr {\em et al.} \cite{Carr-LD01-64pra033603} solved the 1-D NLSE for
a finite square-well finding localized solutions when the 1-D eigenenergy, 
$\varepsilon_1<0$.
The transition at $\varepsilon_1=0$ was found in terms of an approximate
expression for $g_{1,max}$ containing $R_0\sqrt{V_0}$ terms where
the potential well has depth, $V_0$, and width, $2R_0$.
(see Eq.~(21) of Ref.~\cite{Carr-LD01-64pra033603}) which our results
will show agreement with.

Leboeuf and Pavloff \cite{Leboeuf-P01-64pra033602} derived approximate
expressions in terms of the maximum number of atoms that a
1-D potential could support based on the area enclosed by the
potential, $\lambda = |\int_{-\infty}^{\infty} V^\prime_1(x) dx|$.
In the low-density BEC limit, this translates from their $\hbar=m=1$
units to $g_{1,max} = 2 \lambda$ when in o.u..  
They argued that, in general, this was expected 'to be very accurate' 
despite using a $V^\prime_1(x) \to - \lambda \delta(x)$ approximate mapping.
They did not explicitly validate their formula against explicit calculations.
More recent analytical work by Seaman {\em et al.} 
\cite{Seaman-BT05-71pra033609} showed that, for a potential $V^\prime_1(x) 
= -\beta \delta(x)$, the $g_{1,max} = 4 \beta$ exactly.  
Our results agree with Seaman {\em et al.}, in that, there is a
factor of 2 underestimate in the treatment of Leboeuf and Pavloff
in the limit of weak potentials.  Our results also agree with
Carr {\em et al.} \cite{Carr-LD01-64pra033603} which show that
the Leboeuf and Pavloff expression has a factor of 2 overestimate
in the limit of strong potentials.

The final paper of direct relevance to the present study is that
of Adhikari \cite{Adhikari-SK07-42ejpd279} who solved the
3-D NLSE with a finite spherical-well potential,
and computed the maximum nonlinear coupling constant, $g_{3,max}$.
There, a 3-D variational ansatz was applied and an almost linear
relation was plotted (see their Fig.~(3)) in which $g_{3,max} \propto V_0$,
the depth of the well.
Adhikari noted that the variational ansatz underestimates
the magnitude of $g_{3,max}$, but gave no calculations to
quantify this.

Furthermore, of particular interest is that Adhikari 
\cite{Adhikari-SK07-42ejpd279} also applied a Thomas-Fermi approximation 
(TFA) to compute,
in our notation, $g_{3,max} = 4\pi R_0^3 V_0 /3$. 
In their paper it is noted the TFA ``is inadequate for calculating"
$g_{3,max}$, leading to a ``much smaller" value than the variational ansatz.
It was thus surprising to us when we applied the same TFA in
preliminary 1-D calculations \cite{LopezMiranda-JA09-XXIXicpigPA1.3},
and found excellent agreement with explicit calculations of $g_{1,max}$.
In the present paper we find that the TFA gives $g_{2,max}$ and $g_{3,max}$'s
accurately for all of the potential defects that we consider.
Our 3-D results show that both the $g_{3,max}$ results and conclusions
of Adhikari's variational ansatz  \cite{Adhikari-SK07-42ejpd279} are not accurate.




In this work, we present some analytical results in Sec. \ref{sec:analytical} 
using the TFA for $i=$ 1-, 2-, and 3-D defect potentials 
and deduce universal scaling rule properties for $g_{i,max}$ and the 
wavefunction (density profiles).
The $g_{i,max}$ is seen to follow a scaling law that depends on $R_0\sqrt{V_0}$,
across a wide range of width parameters and potential shapes.
In the 1-D case, we present results for a square-well, triangle, truncated 
harmonic, truncated double harmonic, and truncated half circle defect.
We have previously published results of four of these trap shapes
solely in 1-D for $g_{1,max}$ \cite{LopezMiranda-JA09-XXIXicpigPA1.3}.  
Here, we expand and extend that
work into 2-D and then 3-D for a wider range of potential defects.
For the 2-D traps case, we study a rectangle well, truncated harmonic
and pyramid defect.
For a 3-D trap we use a spherical, rectangular cuboid and a truncated harmonic
well defect.
Furthermore, we report a scaling rule for the wavefunction in terms of the
range of the potential.
In Sec. \ref{sec:numerical}, we present our numerical approach to solve
the NLSE based on a finite-difference method on a numerical lattice
and a Gauss-Seidel procedure to obtain the numerical ground state
on the lattice for the 1-, 2-, and 3-D NLSE.
In Sec. \ref{sec:results}, we present and discuss our results.
Finally, in Sec. \ref{sec:conclusions} we present our conclusions.
 
\section{Analytic results using the Thomas-Fermi approximation}
\label{sec:analytical}

The TFA considers the limit of strong interactions between atoms that
form an ultra-cold BEC and allows for some useful expressions for
the single-particle wavefunction to be obtained
\cite{Edwards-M95-51pra1382,Baym-G96-76prl6}.
A BEC is said to be in the Thomas-Fermi (TF) regime when the interaction 
energy dominates over the zero-point energy 
\cite{Edwards-M95-51pra1382,Baym-G96-76prl6}.
The TF states are strictly localized by a potential, and thus their behavior
was not initially expected to be able to mimic the NLSE localized states as 
they approach delocalization as was previously noted by
Adhikari \cite{Adhikari-SK07-42ejpd279}.  
Instead, we will show that this approximation gives us a general 
($R_0\sqrt{V_0}$)-based scaling law that agrees with the numerical NLSE 
solutions.

The TFA neglects the kinetic energy in the NLSE, and therefore the 
time-independent NLSE in either $i \in 1,2,3$ dimensions becomes
\begin{equation}
\label{eqn:tf}
\left[V^\prime_i(\mathbf{r})+g_i^{TF}\vert\Psi_{TF}(\mathbf{r})\vert^2\right]\Psi_{TF}(\mathbf{r})=
\varepsilon_i \Psi_{TF}(\mathbf{r}),
\end{equation}
where $\varepsilon_i$ is the $i$-D single-particle orbital energy.  
The maximum nonlinear coupling constant due to the defect potential
occurs when $\varepsilon_i \to 0$, so Eq.~(\ref{eqn:tf}) reduces to
\begin{equation}
\label{eqn:psitf}
\Psi_{TF}(\mathbf{r}) = \lim_{\varepsilon_i \to 0}\sqrt{\frac{\varepsilon-V^\prime_i(\mathbf{r})}{g_i^{TF}}}
                   = \sqrt{\frac{-V^\prime_i(\mathbf{r})}{g^{TF}_{i,max}}}.
\end{equation}
Through normalization, 
$\int_{\Omega} \vert\Psi(\mathbf{r})\vert^2 dr^i = 1$, one obtains
that $g^{TF}_{i,max}$ satisfies the relation
\begin{equation}
g^{TF}_{i,max}=-\int_{\Omega} V^\prime_i(\mathbf{r}) dr^i \,,
\label{eqn:gTFmax}
\end{equation}
where $\Omega$ is the contour around the defect where 
$V^\prime_i(\mathbf{r})=0$.
Thus, the non-linear coupling parameter $g^{TF}_{i,max}$ just depends on the 
extension $\Omega$ (area or volume for the 2- or 3-D case) of the defect 
potential in the TFA.
Hence, for a given scattering length $a_{s}$, atom mass $m$ and transverse
frequency $\omega_{\perp}$, the corresponding maximum number
of atoms trapped by a defect, $N_{max}$, can be determined.

\subsection{One-dimensional case}

We consider here the following five 1-D defect potential shapes.  
In all of these cases, $V_0>0$ is the strength of the potential and $2R_0$ 
is the width of the potential region.
A square potential
\begin{eqnarray}
V_{1,s}^\prime(x) = \left\{
\begin{array}{cl}
-V_{0} &, |x| < R_{0},\\
0 &, |x| > R_{0},\\
\end{array}\right.
\label{eqn:Vw}
\end{eqnarray}
a half circle potential trap
\begin{eqnarray}
V_{1,c}^\prime(x) = \left\{
\begin{array}{cc}
-V_{0}\sqrt{1-\left(\frac{x}{R_{0}}\right)^{2}} &, |x|<R_{0},\\ 
0 &, |x| > R_{0}.\\
\end{array}\right.
\label{eqn:Vc}
\end{eqnarray}
a (truncated) harmonic potential trap 
\begin{eqnarray}
V_{1,h}^\prime(x) = \left\{
\begin{array}{cc}
-V_{0}\left[\left(\frac{x}{R_{0}}\right)^{2}-1\right] &, |x| < R_{0},\\ 
0 &, |x| > R_{0},\\
\end{array}\right.
\label{eqn:Vh}
\end{eqnarray}
a symmetric double harmonic potential trap
\begin{eqnarray}
V_{1,2h}^\prime(x) = \left\{
\begin{array}{cc}
4V_{0}\left(\left(\frac{x}{R_{0}}\right)^{2}+\frac{x}{R_{0}}\right) &,
-R_{0}<x<0,\\ 
4V_{0}\left(\left(\frac{x}{R_{0}}\right)^{2}-\frac{x}{R_{0}}\right) &,
0<x<R_{0},\\ 
0 &, |x| > R_{0},\\
\end{array}\right.
\label{eqn:V2h}
\end{eqnarray}
and a triangle potential
\begin{eqnarray}
V_{1,t}^\prime(x) = \left\{
\begin{array}{cc}
-V_{0}\left(1+\frac{x}{R_{0}}\right) &, -R_{0}<x<0,\\ 
-V_{0}\left(1-\frac{x}{R_{0}}\right) &, 0<x<R_{0},\\
0 &, |x|> R_{0},\\
\end{array}\right.
\label{eqn:Vt}
\end{eqnarray}

From Eq.~(\ref{eqn:gTFmax}), each potential gives, respectively, a non-linear 
coupling constant given by
\begin{eqnarray}
\label{eqn:gTFmaxav}
g^{TF}_{1,max} = \left\{
\begin{array}{cl}
  2 (R_0\sqrt{V_{0}})^{2}/R_0 , & \mathrm{square-well}\\
  \pi (R_0\sqrt{V_{0}})^{2}/(2R_0),  & \mathrm{circle} \\
  4(R_0\sqrt{V_{0}})^{2}/(3R_0) , & \mathrm{harmonic} \\
  4(R_0\sqrt{V_{0}})^{2}/(3R_0) , & \mathrm{double} \\
  (R_0\sqrt{V_{0}})^{2}/R_0  , & \mathrm{triangle} \,.
\end{array}\right.
\end{eqnarray}
Thus, it seems that $g_{1,max}^{TF}$ has a universal dependence on 
$R_0\sqrt{V_0}$ in the 1-D case, independent of the shape of the defect 
potential, {\em i.e} $g_i=f(R_0\sqrt{V_0})$.

A more general, but still approximate, expression for the nonlinear
coupling constant that includes the kinetic energy term in the NLSE has 
been found by Carr {\em et al.} for the one-dimensional square-well, 
(see Eq.~(21) in Ref.~\cite{Carr-LD01-64pra033603}), and is given by
\begin{equation}
  \label{eqn:gcarr}
g_{1,max}(\gamma)\!\approx\!\frac{\gamma}{R_{0}}\!\left[\sqrt{2}\left(\frac{
e^{-\sqrt{2}\gamma}}{1+2e^{-\sqrt{2}\gamma}-e^{-2\sqrt{2}\gamma}}\right)\!+
\!2\gamma\right],
\end{equation}
where $\gamma = R_{0}\sqrt{V_{0}}$. 
In the limit when $\gamma \gg 1$, the exponentials can be neglected
and Eq.~(\ref{eqn:gcarr}) reduces to $g_{1,max}(\gamma)=2(R_{0}
\sqrt{V_{0}})^{2}/R_{0}$.
This agrees with the TFA found for the case of the square-well
(first line in Eq.~(\ref{eqn:gTFmaxav})). 
The other defect potentials considered here appear to have no equivalent
analytical expression in the literature.

\subsection{Scaling of the 1-D NLSE}

Due to the chosen geometries of the defects, one can make an additional 
change of variables to $\bar{x}=x/R_0$ and $\bar{\varepsilon}_1=\varepsilon_1
/V_0$ obtaining the following reduced 1-D NLSE,
\begin{equation}
\left[-\frac{1}{2V_0R_0^2}\frac{\partial^2\:}{\partial
\bar{x}^2}+\bar{V}(\bar{x})+\frac{g_1 R_0}{V_0R_0^2}\vert\Phi\vert^2\right]\Phi(\bar{x})=
\bar{\varepsilon}_1 \Phi(\bar{x}),
\label{eqn:GPEred}
\end{equation}
where $\Phi(\bar{x})=\sqrt{R_0}\Psi(\bar{x})$ such that $\Phi(\bar{x})$
remains normalized in the $\bar{x}$ range.
Here, $\bar{V}(\bar{x})=V^\prime_1(\bar{x})/V_0$ is the reduced defect potential
with a maximum strength of $-1$ and defined only in the range $|\bar{x}|<1$.

For large values of $(R_0\sqrt{V_0})^2$, one notices that the kinetic energy 
term in Eq.~(\ref{eqn:GPEred}) can be neglected, justifying the
TFA when $R_0$ and/or $V_0$ are large.
The non-linear term can not be neglected since $g_{1}=f(R_0\sqrt{V_0})$,
as previously mentioned.
The equivalent of Eq.~(\ref{eqn:gTFmax}) in these reduced units can be 
thought of as a shape factor, $\alpha$, given by the area under the 
reduced potential:
\begin{equation}
\alpha_{1D} \equiv \frac{g_{1,max}^{TF}R_0}{(R_0\sqrt{V_0})^2}
                 = -\int_{|\bar{x}|<1} \bar{V}(\bar{x})d\bar{x}
\label{eqn:formfactor}
\end{equation}
This shape factor takes the values of $2,\,\pi/2,\,4/3,\,4/3$, and 
$1$ for each 1-D potential, respectively.

On the other hand, from Eq.~(\ref{eqn:GPEred}), we now note that 
defects of the same type with the same $R_0\sqrt{V_0}$ have the same 
$g_{1,max}$ and the same reduced wavefunction, therefore, there is a   
fundamental relation between the density profile for different parameters.
We will discuss below some further examples where this scaling law holds.

\subsection{Two-dimensional case}

For the case of a two-dimensional traps, the time-independent NLSE is given by
\begin{eqnarray}
\Big\{-\frac{1}{2}\left[\frac{\partial^2}{\partial x^2} +
\frac{\partial^2}{\partial y^2}\right]+V^\prime_2(x,y)&+&  \\
g_2|\Psi|^2 \Big\}\Psi(x,y) &=&\varepsilon_2 \Psi(x,y) \,.\nonumber
\end{eqnarray}
Thus, the application of the TFA again gives the following expression 
for the 2-D coupling constant at delocalization
\begin{equation}
  g_{2,max}^{TF} = -\int_{\Omega}V^\prime_2(x,y)\,dx dy \,.
\end{equation}
In this case we have considered the following three 2-D
defect potentials: a rectangular well
\begin{equation}
V^\prime_2(x,y)=\left\{\begin{array}{cc}
-V_0 &, \mathrm{if}\; |x| <R_x\; \mathrm{and}\; |y|<R_y \\
0 &, \mathrm{if} \;|x| >R_x \;\mathrm{and}\; |y|>R_y\,, \\
\end{array}\right.
\end{equation}
a truncated harmonic (parabolic) well with circular base
\begin{equation}
V^\prime_2(x,y)=\left\{\begin{array}{cc}
-V_0(R_0^2-x^2-y^2)/R_0^2 &, \mathrm{if}\; \sqrt{x^2+y^2} < R_0 \\
0 &, \mathrm{otherwise}\,,  \\
\end{array}\right.
\end{equation}
thirdly, a pyramid (triangle) potential 
\begin{equation}
V^\prime_2(x,y)=\left\{\begin{array}{rr}
-V_0\frac{(y+R_y)}{R_0} &,  y< |x|\,,\, -R_y<y<0 \\
 V_0\frac{(y-R_y)}{R_0} &,  y> |x|\,,\, R_y>y>0 \\
-V_0\frac{(x+R_x)}{R_0} &,  |y|< x\,,\, -R_x<x<0 \\
 V_0\frac{(x-R_x)}{R_0} &,  |y|> x\,,\, R_x>x>0 \\
 0 &, \mathrm{otherwise} \\
\end{array}\right.
\end{equation}
where each line represents the equation on each pyramid wall.

\subsection{Scaling of the 2-D NLSE}

Following a procedure similar to the 1-D case, the
2-D NLSE can be further rewritten in terms of
the reduced variables.  Making the change of variables:
$\bar{x}=x/\sqrt{R_xR_y}$, $\bar{y}=y/\sqrt{R_xR_y}$,
$\Phi=\sqrt{R_xR_y}\Psi$, and $\bar{\varepsilon}_2=\varepsilon_2/V_0$, 
one obtains
\begin{eqnarray}
\label{eq:gpe-2d}
\Big\{-\frac{1}{2V_0R_xR_y}\left[\frac{\partial^2}{\partial \bar{x}^2} +
\frac{\partial^2}{\partial \bar{y}^2}\right]+\bar{V}_2^\prime(\bar{x},\bar{y})&+&
\\
\frac{g_2}{V_0R_xR_y}|\Phi|^2 \Big\}\Phi(\bar{x},\bar{y}) &=&\bar{\varepsilon}_2
\Phi(\bar{x},\bar{y}) \nonumber
\end{eqnarray}
%
where $\bar{V}_2 = V_2^\prime/V_0$ is the 2-D reduced potential.
For example, for the rectangular well, we have
\begin{equation}
\bar{V}_2(\bar{x},\bar{y})=\left\{\begin{array}{cc}
-1 &, \mathrm{if}\; |\bar{x}| <\sqrt{\frac{R_x}{R_y}}\; \mathrm{and}\;
|\bar{y}|<\sqrt{\frac{R_y}{R_x}} \\
0 &, \mathrm{otherwise}\,. \\
\end{array}\right.
\end{equation}

This gives the following scaling law and expressions for the nonlinear 
2-D coupling constant in terms of the 2-D shape factor
\begin{equation}
\alpha_{2D} = \frac{g_{2,max}^{TF}}{V_0R_xR_y}
            =-\int\int \bar{V}_2(\bar{x},\bar{y})d\bar{x}d\bar{y}
\end{equation}
where the limits on integration are for $|\bar{x}|<\sqrt{R_x/R_y}$ and 
$|\bar{y}|<\sqrt{R_y/R_x}$. 
For the defects considered here,
\begin{equation}
\label{eq:g-2d-tf}
g_{2,max}^{TF}=\left\{\begin{array}{cl}
4V_0R_xR_y &, \mathrm{Rectangle} \\
\frac{\pi}{2}V_0R_0^2 &, \mathrm{Parabolic} \\
\frac{4}{3}V_0R_xR_y &, \mathrm{Pyramid}\,.\\
\end{array}\right.
\end{equation}
such that the three shape factors are $4$, $\pi/2$, and $4/3$ and thus
$g_2=f(V_0R_xR_y)$ which for a symmetric case becomes
$g_2=f(R_0\sqrt{V_0})$.
Again, large values of $V_0R_xR_y$ can justify the TFA since the kinetic
energy term can be neglected in Eq. (\ref{eq:gpe-2d}), but not the 
non-linear term.

\subsection{Three-dimensional case}

For the case of three-dimensional traps, the time-independent NLSE in
oscillator units is given by
\begin{eqnarray}
\Big\{-\frac{1}{2}\left[\frac{\partial^2}{\partial x^2} +
\frac{\partial^2}{\partial y^2}+\frac{\partial^2}{\partial z^2}\right]
    +V^\prime_3(x,y,z)+  \\ g_3|\Psi|^2 \Big\}\Psi(x,y,z) =\varepsilon_3
\Psi(x,y,z) \,.\nonumber
\end{eqnarray}
In this case we have considered three different 3-dimensional
defect potentials. Firstly, a spherical well,
\begin{equation}
V^\prime_3(x,y,z)=\left\{\begin{array}{cc}
-V_0 &, \mathrm{if}\; r <R_0 \\
0 &, \mathrm{if} \;r >R_0
\end{array}\right.
\end{equation}
where $r=\sqrt{x^2+y^2+z^2}$.
Secondly, a rectangular cuboid well,
\begin{equation}
V^\prime_3(x,y,z)=\left\{\begin{array}{cc}
-V_0 &, \mathrm{if}\; |x| <R_x, \; |y|<R_y, \\
  &  \mathrm{and}\; |z|<R_z \\
0 &, \mathrm{if} \;|x| >R_x, \; |y|<R_y, \\
  & \mathrm{and}\; |z|>R_z\,, \\
\end{array}\right\}
\end{equation}
(which forms a cube when $R_x=R_y=R_z$).
Thirdly, a truncated harmonic (parabolic) well,
\begin{equation}
V^\prime_3(x,y,z)=\left\{\begin{array}{cc}
-V_0(1-r^2/R_0^2) &, \mathrm{if}\; r < R_0 \\
0 &, \mathrm{otherwise}\,,  \\
\end{array}\right.
\end{equation}
where $r=\sqrt{x^2+y^2+z^2}$.

The application of the TFA again gives the expression for the 
three-dimensional constant,
\begin{equation}
  g_{3,max}^{TF} = -\int_{\Omega}V^\prime_3(x,y,z)\,dx dy dz\,.
\end{equation}

\subsection{Scaling of the 3-D NLSE}

Repeating the procedure used in the 1- and 2-D case, the 3-D NLSE can 
be rewritten in terms of reduced variables.  
Making the changes,
$\bar{x}=x/(R_xR_yR_z)^{1/3}$, $\bar{y}=y/(R_xR_yR_z)^{1/3}$, and
$\bar{z}=z/(R_xR_yR_z)^{1/3}$ for the reduced coordinates and 
$\Phi=\sqrt{R_xR_yR_z}\Psi$ for the reduced wavefunction and 
$\bar{\varepsilon}_3=\varepsilon_3/V_0$ for the energy, one obtains
\begin{eqnarray}
\label{eq:gpe-3d}
\Big\{-\frac{1}{2V_0(R_xR_yR_z)^{2/3}}\left[\frac{\partial^2}{\partial 
\bar{x}^2} + \frac{\partial^2}{\partial \bar{y}^2}+\frac{\partial^2}{\partial 
\bar{z}^2}\right]&  & \\ 
+\bar{V}_3(\bar{x},\bar{y},\bar{z})+\frac{g_3/(R_xR_yR_z)^{1/3}}
{V_0(R_xR_yR_z)^{2/3}}|\Phi|^2 \Big\} &&
\Phi(\bar{x},\bar{y},\bar{z})  \nonumber \\
=\bar{\varepsilon}_3 &&\Phi(\bar{x},\bar{y},\bar{z})  
\nonumber
\end{eqnarray}
where $\bar{V}_3 = V^\prime_3/V_0$ is the 3-D reduced defect potential.

For example, for the rectangular cuboid well, we have
\begin{equation}
\bar{V}_3(\bar{x},\bar{y},\bar{z})=\left\{\begin{array}{cc}
-1 &, \mathrm{if}\; |\bar{x}| <\left(\frac{R_x^2}{R_yR_z}\right)^{1/3}, \\
 & |\bar{y}|<\left(\frac{R_y^2}{R_xR_z}\right)^{1/3}, \\
 & \mathrm{and}\; |\bar{z}|<\left(\frac{R_z^2}{R_xR_y}\right)^{1/3} \\
0 &, \mathrm{otherwise}\,. \\
\end{array}\right.
\end{equation}

In this case, the TFA gives the following scaling law and expressions for 
the nonlinear 3-D coupling constant in terms of the 3-D shape factor
\begin{equation}
  \alpha_{3D} = \frac{g_{3,max}^{TF}}{V_0R_xR_yR_z}
              =-\int\int\int \bar{V}_3(\bar{x},\bar{y},\bar{z})d\bar{x}d\bar{y}d\bar{y}\,,
\end{equation}
where the integration is over the range $|\bar{x}|<(R_x^2/R_yR_z)^{1/3}$, 
$|\bar{y}|<(R_y^2/R_xR_z)^{1/3}$, and $|\bar{z}|<(R_z^2/R_xR_y)^{1/3}$.

For the case of a spherical well, a rectangular cuboid, and harmonic 
parabolic well, we have that the TFA gives
\begin{equation}
\label{eq:g-3d-tf}
g_{3,max}^{TF}=\left\{\begin{array}{cl}
4\pi V_0R_0^3/3 &, \mathrm{Spherical} \\
8V_0R_xR_yR_z &, \mathrm{Cuboid} \\
\frac{8\pi}{15}V_0R_0^3 &, \mathrm{Parabolic}\,, \\
\end{array}\right.
\end{equation}
such that the shape factor is $4\pi/3$, 8 and $8\pi/15$, respectively.
Similarly to the 1- and 2-D case, $g_3=f(V_0R_xR_yR_z)$, which for a
symmetric case $g_{3,max}/R_0=f(R_0\sqrt{V_0})$, such that
when large values of $V_0(R_xR_yR_z)^{2/3}$ are involved, then
the TFA is justified in Eq. (\ref{eq:gpe-3d}). Remember, the non-linear term 
can not be neglected.

Thus, in summary, defining the defect shape factor
\begin{equation}
\label{eq:shape}
\alpha_i=-\int_{\Omega} \bar{V}_i(\mathbf{r})d^i\mathbf{r} \: ,
\end{equation}
where $i$ is the dimension of the space and the integration is on the
$\Omega$ space that contains the trap. Then the maximum nonlinear coupling
constant that a defect can support is given by
\begin{equation}
g_{i,max}^{TF}=\left\{\begin{array}{lc}
\alpha_{i}V_0R_x, & i=\rm{1D}\\
\alpha_{i}V_0R_xR_y, & i=\rm{2D}\\
\alpha_{i}V_0R_xR_yR_z, & i=\rm{3D}\,,\\
\end{array}\right.
\end{equation}
and, for the particular case of symmetric traps where $R_x=R_y=R_z=R_0$, then
\begin{equation}
\label{eq:galpha}
g_{i,max}^{TF}=\left\{\begin{array}{lc}
\alpha_{i}(\sqrt{V_0}R_0)^2/R_0, & i=\rm{1D}\\
\alpha_{i}(\sqrt{V_0}R_0)^2, & i=\rm{2D}\\
\alpha_{i}(\sqrt{V_0}R_0)^2R_0, & i=\rm{3D}\,.\\
\end{array}\right.
\end{equation}
That is, 
\begin{equation}
g_{i,max}R^{i-2}=f(R_0\sqrt{V_0})
\end{equation}
for the $i$-dimension case and the larger the shape factor, the larger the 
number of trapped atoms by the well.

In order to verify these expressions outside of the TFA, let us solve
Eqs.~(\ref{eqn:GPEred}), (\ref{eq:gpe-2d}), and (\ref{eq:gpe-3d}) for the
reduced wavefunction, $\Phi$, by a numerical procedure.
In the next section we will outline the numerical method we implemented to
solve the NLSE for one, two, and three-dimensions for solutions approaching
the point of wavefunction delocalization, that is, when $\varepsilon_i \to 0$.

\section{Numerical approach}
\label{sec:numerical}

In this section we present the two computational methods we used to
compute the wavefunction, and the algorithm we used to determine the $g_{max}$.

\subsection{Crank-Nicolson method}

By using finite-differences and the Crank-Nicolson method (CN)
\cite{CNmetodo,NumRecipes,Goldberg-A67-35ajp177}, 
the ground state solutions to the time-independent NLSE can be found for a
given non-linear coupling constant $g$ without approximation.
To do so, the time-dependent NLSE is evolved in negative imaginary time 
\cite{Esry-BD97-55prl3594}.
The kinetic and potential energy terms can be efficiently evolved by means
of a symmetric split-operator method \cite{Bromley-MWJ04-69pra053620}
\begin{equation}
  \Phi(\bar{\mathbf{r}},t_0+\Delta t)
  \approx e^{-i\Delta t \mathbf{\hat{V}}/2} \:
          e^{-i\Delta t \mathbf{\hat{T}}} \:
          e^{-i\Delta t \mathbf{\hat{V}}/2} \: \Phi(\bar{\mathbf{r}},t_0).
\end{equation}
Here $\mathbf{\hat{T}}$ is the kinetic energy operator and
$\mathbf{\hat{V}}$ is the potential energy operator in the NLSE.

This requires that the wavefunction to be discretized in space on a
numerical grid, {\em viz.} $\Phi(\bar{x}_i,\bar{y}_j,\bar{z}_k,t_n) \rightarrow \Phi_{ijk}^n$.
The 1-D NLSE in this approach, for example, becomes
\begin{eqnarray}
\label{eqn:nlsdiscrete}
\lefteqn{ \lbrace\xi_k-\nu(\xi_{k+1}-2\xi_{k}+
\xi_{k-1})\rbrace= } \nonumber\\
& & \lbrace f_{k}+\nu(f_{k+1}-2f_{k}+f_{k-1})\rbrace \; ,
\end{eqnarray}
where $\nu=i\Delta t/(4\Delta \bar{x}^2 V_0R_0^2)$, and the potentials are in 
$f_k=\exp\{-i \Delta t V_k^n/2\} \Phi_k^n$ and
$\xi_k=\exp\{i \Delta t V_k^{n+1}/2\} \Phi_k^{n+1}$ with
$V_k^n=\bar{V}_k+g_1R_0/(V_0R_0^2)|\Phi_k^n|^2$.
Eq. (\ref{eqn:nlsdiscrete}) can be written in matrix form as
$\mathbf{A^{+}}\overrightarrow{\Phi}^{n+1}=\mathbf{A^{-}}\overrightarrow{f}$.
%
%
Note that $\textbf{A}^{\pm}$ is a constant matrix for fixed $\Delta t$
and $\Delta \bar{x}$.
For the multi-dimensional problem, the unitary operators are applied
in sequence, {\em e.g.} $e^{-i\Delta t \mathbf{\hat{T}}} = e^{-i\Delta t 
\mathbf{\hat{T}_x}} e^{-i\Delta t \mathbf{\hat{T}_y}} e^{-i\Delta t 
\mathbf{\hat{T}_z}}$, thus only requiring (many) tri-diagonal matrix solves 
at intermediate stages.

%

\subsection{Gauss-Seidel method}

We also implemented the Gauss-Seidel (GS) method \cite{Kooning} in 1-D, 
2-D and 3-D to find the ground state solution of the time-independent NLSE.  
In this case, the energy is evaluated given an {\em improved} solution at
each point of the numerical grid.  
The GS is much simpler than the CN method, and serves as a valuable 
cross-check that our solutions are converged.

For example, in the 1-D case, the wavefunction evaluated in the $k$-th 
grid point, $\Phi_k$, is replaced by $\Phi_k^\prime$, where
\begin{eqnarray}
   \Phi_k^\prime & \approx & (1-\beta)\Phi_k + \\
& & \frac{\beta \left(\Phi_{k+1}+\Phi_{k-1}\right)}{2[1+(\bar{V}_kV_0+
g_i\Phi_k^2/R_0-\varepsilon_i) (R_0\Delta \bar{x})^2)]}\,, \nonumber 
\label{eqn:GS}
\end{eqnarray}
%
where $\beta$ is the relaxation parameter that ensures convergence
to the lowest energy state.  In our case we take $\beta=3/4$
as a compromise between convergence time and precision \cite{Kooning}.
Extension of the GS method to 2-D and 3-D is straightforward.

\begin{figure}[!t]
 \centering
 \includegraphics[width=240pt]{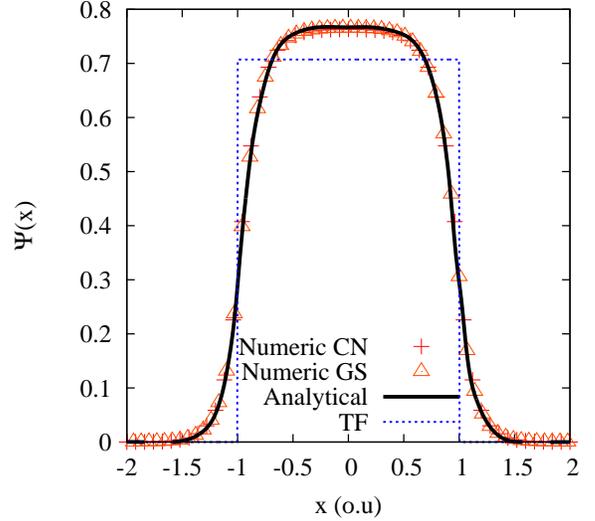}
 \caption{(Color on-line). The computed NLSE ground state wavefunction
for a 1-D square-well defect.
The ($+$) symbols are the results of the CN method and
($\bigtriangleup$) for the GS method in a square-well with
$g_1=25$, $V_{0}=50$, and $R_0=1.0$ o.u. which gives $\varepsilon_1=-35.41668$ 
o.u..
The analytical wavefunction solution (solid line) of Ref.
\cite{Carr-LD01-64pra033603}, and the TFA (dashed line) of
Eq.~(\ref{eqn:psitf}) are also shown for comparison.}
 \label{fig:fo}
\end{figure}

The results of both the GS and CN computational methods are illustrated
in Fig.~\ref{fig:fo}. 
This shows the wavefunction, $\Psi(x)$, for a 1-D NLSE with a strong 
non-linearity ($g_1=25$ and $\varepsilon_1 =-35.41668$),
that is trapped by a strong square-well ($V_{0}=50$ and $R_0=1.0$ o.u.).
Note that these are all given in oscillator units, and not the reduced units.
This is compared with the analytical wavefunction of Carr \textit{et al.}
(see Eq.~(12) in Ref.~\cite{Carr-LD01-64pra033603}).
The excellent agreement validates our numerical results.
Also shown is the TFA of Eq.~(\ref{eqn:psitf}), where the
wavefunction takes the same shape as the defect and is identically zero
outside the well.  

\subsection{Determination of $g_{i,max}$}
\label{sec:gmaxdeterm}

The main computational challenge is to systematically increase $g$ to 
find $g_{i,max}$ for a given $V_{0}$ and $R_0$ for $i=1$-, 2-, and 3-D
potentials.  
Thus, the problem is reduced to a root search for which
$\varepsilon_i(g_{i,max})=0$ for a given $R_0$ and $V_0$.
As $g_i$ is increased, the wavefunction penetrates further into the 
classically forbidden region, until $g_{i,max}$ is reached, at that 
point the solution determined is no longer a localized wavefunction
\cite{Carr-LD01-64pra033603}.
To determine $g_{i,max}$ (and hence $N_{max}$) numerically, it is required
that the discretized grid be large enough to contain the very slow 
decay of the wavefunction when $g_i$ approaches $g_{i,max}$.  
Note, however, that
we strictly use $\Phi(\pm \bar{x}_{max}) = 0$, where we effectively have a
box $\bar{V}_1(\pm \bar{x}_{max}) \to \infty$.  
This means that our $\varepsilon_i >0$ scattering states are artificially 
constrained and are always square integrable.

To ensure that the determined $g_{i,max}$ is accurate to the precision that
we demand, the results simply need to be insensitive to the location of
the boundary.  
In our 1-D cases we could simply choose $\bar{x}_{max}= 300$ and 
$\Delta \bar{x}=0.01$.  
The $g_{i,max}$'s were determined by choosing the defect parameters $V_0$ 
ranging from 0.01, 0.1, 0.5, 1.0, 5.0, 10.0, 20, and 50.0 o.u., while for 
the width we choose $R_0$ from 0.5, 1.0, 2.0, 5.0, 10.0, and 20.0 o.u. 
commensurate with physical parameters of typical BEC experiments (as 
discussed later in Sec. \ref{sec:app}).

For the 2-D cases we chose $\bar{x}_{max}=\bar{y}_{max}=5.0$ in the
reduced units space where the potential reaches only unity range and
$\Delta \bar{x}=\Delta \bar{y}=0.05$ which gives a 200$\times$200 grid
points in the wavefunction.
For the 3-D cases, we similarly used $\bar{x}_{max}=\bar{y}_{max}=
\bar{z}_{max}=5.0$ and $\Delta \bar{x}=\Delta \bar{y}=\Delta \bar{z}=0.05$ 
with a wavefunction size of $200\times 200\times 200$.
Thus, increasing the dimension in the calculation introduces a large memory
footprint, as well as many tri-diagonal matrices to solve in the CN 
calculations.
Compared to the 1-D calculations, we restrict the range of defect parameters
explored to ensure accuracy, whilst still spanning the parameter regimes 
of interest.

In all cases, we assumed convergence for a given $g_i$ when $\varepsilon_i$ 
from one iteration to the next changes less than $\Delta \varepsilon_i=
10^{-8}$.  
The $g_i$ was then incremented until $\varepsilon_i = 0$ was located to 
within $10^{-8}$.
The amount of increment of $g_i$ was 1\% of the TFA initial value. 
When $\varepsilon_i > 0$ was obtained for a given $g_i$, a step back was
performed and $\Delta g_i$ was reduced in half, until we reached
$\varepsilon_i = 0$ from below.
%

\section{Results and Scaling Laws}
\label{sec:results}

\subsection{The 1-D square well case}

The numerical and analytical results for the 1-D square well are shown
in Table~\ref{tab:1dcomparisons} for the smallest and largest value of
$R_0$ used in this work.  
The $g_{1,max}$ value was firstly determined for a range of defects spanning 
large and small values of the product $R_0\sqrt{V_0}$.
We have chosen to translate this into the maximum number of trapped atoms that
this corresponds to, specifically for a cloud of $^{87}$Rb atoms (see caption).
These are then compared against the expressions from the TFA, and those
obtained by Carr \textit{et al.} \cite{Carr-LD01-64pra033603} and  Leboeuf 
and Pavloff \cite{Leboeuf-P01-64pra033602}.

\begin{table}
\caption{\label{tab:1dcomparisons} The maximum nonlinear coupling constant, 
$g_{1,max}$, for 1-D square well potentials with various widths ($2R_0$) 
and depths ($V_0$) as determined by the 1-D numerical calculations.  
The corresponding maximum number of atoms trapped by the potentials are 
given as the $N_{1,max}$ columns assuming that the atomic species is 
$^{87}$Rb (with $a_s = 100 a_0$, where $a_0$ is the Bohr radius) trapped by 
a transverse frequency of $\omega = 2\pi\times 100$ rad/s (thus $a_{\perp} 
= 2.7$ microns).  
The superscripts denote (a) the numerical calculation, and (b) the present 
TFA from Eq.~(\ref{eqn:gTFmaxav}).
The approximations of (c) Carr \textit{et al.} \cite{Carr-LD01-64pra033603}
and (d) Leboeuf and Pavloff \cite{Leboeuf-P01-64pra033602} are given for 
comparison.}
\begin{ruledtabular}
\begin{tabular}{llccccc}
$R_0$ & $V_0$ & $g_{1,max}^{a}$ & $N_{1,max}^{a}$ & $N_{1,max}^{b}$ & 
$N_{1,max}^{c}$ & $N_{1,max}^{d}$ \\
\hline
0.5   & 0.05  & 0.044346        & 12.321          & 13.764          &  48.597         & 26.528 \\
0.5   & 0.1   & 0.141031        & 37.003          & 26.528          &  73.108         & 52.056 \\
0.5   & 0.5   & 0.852832        & 218.71          & 128.64          &  212.55         & 256.28 \\
0.5   & 1     & 1.635992        & 418.64          & 256.28          &  358.41         & 511.56 \\
10    & 0.5   & 10.679472       & 2727.3          & 2553.8          &  2553.8         & 5106.6 \\
10    & 1     & 21.023785       & 5368.0          & 5106.6          &  5106.6         & 10212 \\
10    & 5     & 102.877445      & 26264           & 25529           & 25529           & 51057 \\
10    & 10    & 204.679689      & 52252           & 51057           & 51057           & 102113 \\
\end{tabular}
\end{ruledtabular}
\end{table}

For large $R_0\sqrt{V_0}$, the potential is strongly binding and
our calculations agree closely with both the TFA and that
of Carr \textit{et al.} \cite{Carr-LD01-64pra033603}.
In this limit, as we increase the $g_1$ towards the $g_{1,max}$ value,
the numerical wavefunction resembles the TF wavefunction until we
approach extremely close to the $g_{1,max}$ and the wavefunction
rapidly delocalizes.  Thus, our numerical method for computing
$g_{1,max}$ is quite accurate in this regime and not affected by
the presence of the boundary conditions.
The formula derived by Leboeuf and Pavloff \cite{Leboeuf-P01-64pra033602},
with weak binding potentials in mind, has a factor of 2 overestimate in
the strong binding limit.

It is also interesting, however, to reconcile the values in
Table~\ref{tab:1dcomparisons} in the small $R_0\sqrt{V_0}$ limit.  
Seaman {\em et al.} \cite{Seaman-BT05-71pra033609} gives a factor
of $g_{1,max} = 4 \lambda$, the same as the Carr \textit{et al.} 
\cite{Carr-LD01-64pra033603}
in the limit of a small perturbation.  That is, the system should approach
that of binding to a delta-function of 'strength' $\lambda=2R_0V_0$.
We expect that our $R_0=0.5$, $V_0=0.05$ numerical calculation should be close to 
$g_{1,max} = 4\times 2R_0V_0 = 0.2$ and not the $g_{1,max}=0.05$ which the TFA estimates.
For this weakly bound case we find that, even with our numerical
wall located at $\bar{x}_{max} = \pm 300$, our $g_{1,max}$ and thus
$N_{1,max}$ are underestimated for the weakest traps.  Essentially, the overall
energy of the system is artificially raised 
due to the system being trapped in a finite sized grid,
and thus the $g_{1,max}$ needed to reach delocalization is significantly
underestimated for small $R_0\sqrt{V_0}$.
Finally, Leboeuf and Pavloff \cite{Leboeuf-P01-64pra033602}
give a factor of 2 underestimate for small perturbations
even though they assumed the limit of a small perturbation
in their derivation.

\subsection{The general 1-D trap cases}

\begin{figure}[!t]
 \centering
 \includegraphics[width=240pt]{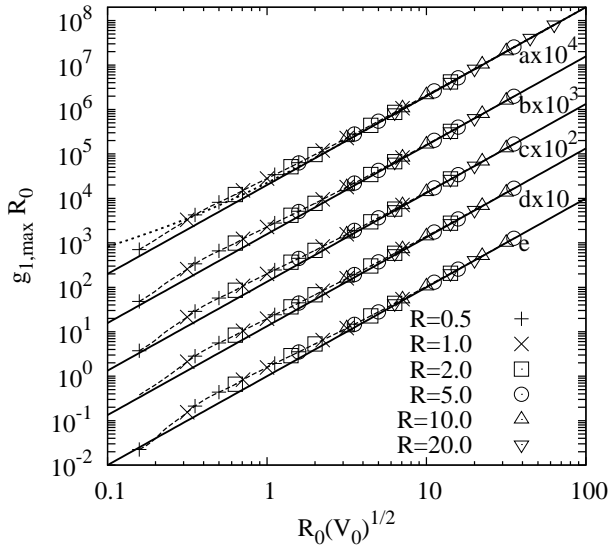}
 \caption{1-D scaling law for $g_{1,max}$ weighted by $R_0$ for a
square (a), circular
(b), harmonic (c), double harmonic (d), and triangular well (e) defect,
respectively, of various lengths $R_0$ and depths $V_0$.
The numerical results are given as various symbols as indicated in
the figure legend, scaled by factors of 10 between each defect shape
to avoid nearly overlapping in the figure.
The long-dashed lines are a guide for the eye along the numerical results.
The TFAs results of Eq.~(\ref{eqn:gTFmaxav}) are the solid lines.
The square-well results are compared with the analytical expression of
Carr \textit{et al.}, Eq.~(\ref{eqn:gcarr}), as the short dashed line.}
 \label{fig:scaledall}
\end{figure}

The results for $g_{1,max}$ for all five 1-D defects are shown in Fig. 
\ref{fig:scaledall} as a function of $R_0\sqrt{V_0}$. 
In order to avoid overlapping, the results for the different shape defects
have been scaled by a factor of 10 between each other in the ascending order 
as given by Eq.~(\ref{eqn:gTFmaxav}).
In the same figure we show the Thomas-Fermi approximations represented by the
solid lines, while the long-dashed line is a guide to the eye that follows the 
numerical results.
For the case of the square-well trap, we also show the approximate analytical 
solution given by Eq. (\ref{eqn:gcarr}) (short-dashed line).

The reason to plot $g_{1,max} R_0$ {\em vs.} $R_0\sqrt{V_0}$ in a log-log scale
and not $g_{1,max}R_0/f(R_0\sqrt{V_0})$ is to show the range of values
that $g_{1,max}$ might take over the range we used in $R_0\sqrt{V_0}$, and
not just to have a constant value defined by the shape factor.

\begin{figure}[!t]
\centering
\includegraphics[width=240pt]{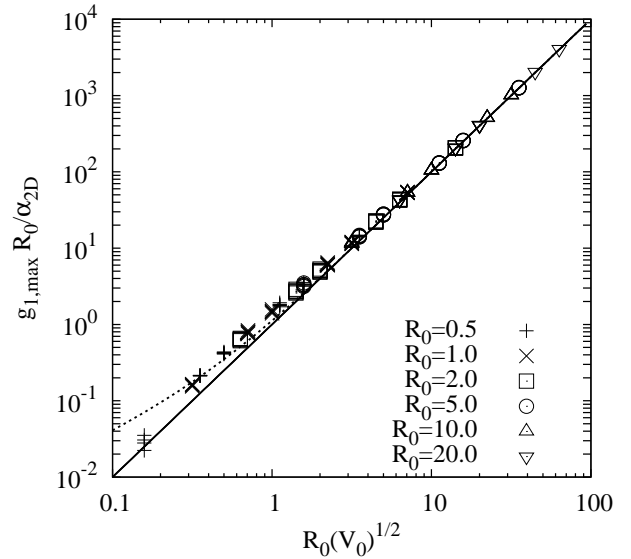}
\caption{Scaling laws for $g_{1,max}R_0/\alpha_{1D}$ as a
function of $R_0\sqrt{V_0}$ for the five different 1-D defect shapes.
The solid line is the TFA which becomes identical for all defects.
The symbols are the numeric results as obtained from
Eq.~(\ref{eqn:GPEred}).
The short-dashed line is the approximation of
Carr {\em et al.} \cite{Carr-LD01-64pra033603} for the square-well.
\label{fig:scaled}
}
\end{figure}

As observed in Fig. \ref{fig:scaledall}, some symbols are seen to overlap, 
confirming
the basic scaling rule for different widths and depths of the same defect.
For large $R_0\sqrt{V_0}$ the numerical results follow closely the TFA.  
This is due to large $R_0\sqrt{V_0}$ which requires a large number of atoms 
to fill up the defect, {\em i.e.}, the system is in a strong nonlinearity 
regime where the TFA is valid.  
Note though, that this also means that our approximation of holding the 
wavefunction in a tight transverse state will eventually breakdown.
For values of $R_0\sqrt{V_0}<1$, the kinetic energy term starts to dominate 
since $g_{1,max}$ gets smaller.  
It is in this region that the difference between the defect shapes is revealed.
However, this is also the region where the 1-D NLSE itself is not valid
anymore as mentioned in the introduction.

As the only difference between the 1-D defects is their shape factors
[Eq.~(\ref{eqn:formfactor})], we can also plot all of them in a single figure. 
Figure \ref{fig:scaled} shows the results by plotting
$g_{1,max}R_0/\alpha_{1D}$ as a function of $R_0\sqrt{V_0}$. 
For $R_0\sqrt{V_0}>1$ there is barely any difference for all of the trapping
defects in the factor $g_{1,max}R_0/\alpha_{1D}$, confirming our generalized 
scaling rule. 
Thus, under these conditions, the square-well defect holds the most atoms. 
It is followed by the circular defect, then by the harmonic well defect and the 
double harmonic well, and finally the triangular defect will hold 
the least atoms for the same width and depth defects.
This is in both the numerical results and the TFA [Eq. (\ref{eqn:gTFmaxav})].

\begin{figure}[!t]
\centering
\includegraphics[width=240pt]{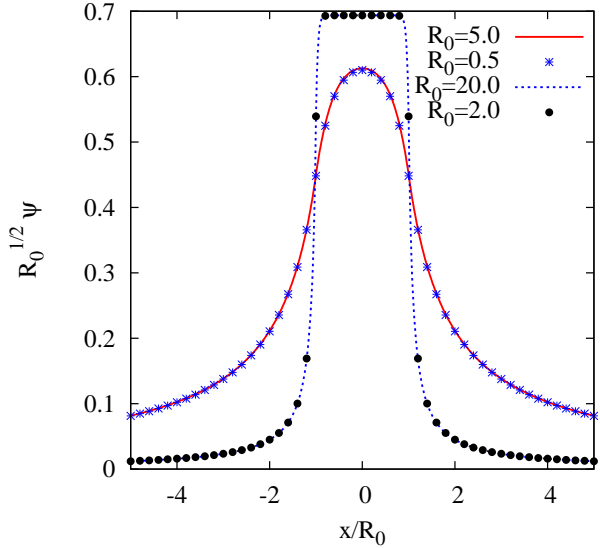}
\caption{(Color on-line).  Scaled 1-D numerical wavefunctions,
$\Phi(\bar{x})=\sqrt{R_0}\Psi(x/R_0)$,
for various square-well defects.
The solid line corresponds to $R_0=5.0$ and $V_0=0.1$ o.u.;
the blue ($*$) symbols correspond to $R_0=0.5$ and $V_0=10.0$ o.u.,
such that both of them have the same $R_0\sqrt{V_0}=1.5811$ and
$g_{1,max}R_0=6.284$.
The dashed line denotes $R_0=20.0$ and $V_0=0.5$ o.u., while the
($\bullet$)
symbol is for $R_0=2.0$ and $V_0=50.0$ o.u., both of these wavefunctions
have
$R_0\sqrt{V_0}=14.1421$ and $g_{1,max}R_0=415.770$, thus, showing the same
reduced wavefunction.
\label{fig:well_wf}
}
\end{figure} 

In Fig. \ref{fig:well_wf}, we show the scaled wavefunctions,
$\Phi(\bar{x}) = \sqrt{R_0} \Psi(\bar{x})$,
for the square-well defects for two values of $g_{1,max}R_0$,
and two values of $R_0\sqrt{V_0}$ as a function of $\bar{x}=x/R_0$. 
The two cases shown give us a small and a large value of $g_{1,max}R_0$.
For the first case, $R_0\sqrt{V_0}=1.5811$ and $g_{1,max}R_0=6.284$, in 
particular, the two 
square-well defect wavefunctions have $R_0=5.0$ and $V_0=0.1$ o.u. (solid 
line) for one case, and $R_0=0.5$ and $V_0=10.0$ o.u. ($*$ symbol) for the 
other case.
That is, a wide and shallow potential trap {\em vs.} a tight and deep potential,
but both with small $R_0\sqrt{V_0}$ value.

We note in Fig. \ref{fig:well_wf} that the scaled wavefunction show the
same tunneling for both cases of $R_0\sqrt{V_0}$.
However, due to the factor $\sqrt{R_0}$ in front of $\Psi$, both cases
would have different tunneling in the $x$ oscillator units space.
The second case considered in Fig.~\ref{fig:well_wf}, shows
the wavefunction of two square-well defects with $R_0\sqrt{V_0}=14.1421$
and $g_{1,max}R_0=415.770$,
in particular, $R_0=20.0$ and $V_0=0.5$ o.u. (dashed line) and
$R_0=2.0$ and $V_0=50.0$ o.u. ($\bullet$ symbol).
Again, we have a wide and shallow potential trap {\em vs.} a tight and deep
trap, but now with a large values for $R_0\sqrt{V_0}$.
In this case we are in the high $R_0\sqrt{V_0}$ region where we are closer 
to the TFA as shown by the shape of the wavefunction which is 
closer to the potential shape. 

\begin{figure}[!t]
\centerline{
\includegraphics[width=240pt]{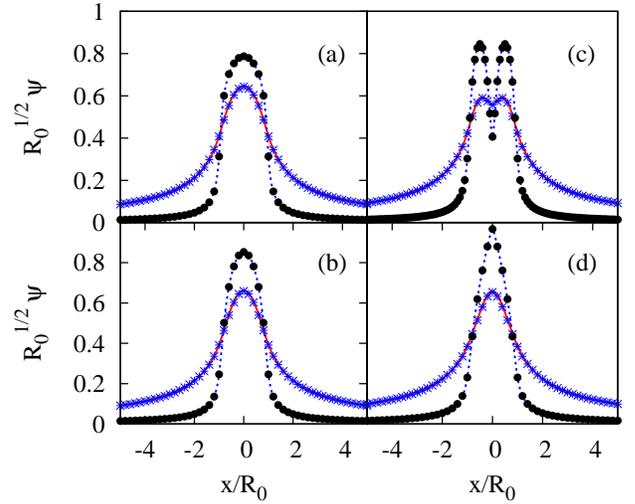}}
\caption{(Color on-line). Scaled 1-D numerical wavefunctions, $\Phi(\bar{x})$,
for the (a) circular, (b) harmonic, (c) double harmonic and (d) triangle 
potential.
The labels and parameters are the same as in Fig. \ref{fig:well_wf}.
\label{fig:wfothers} }
\end{figure}

The scaling was also verified for the other potentials.
In Fig. \ref{fig:wfothers}, we show the corresponding scaled wavefunctions 
for the circular (a), harmonic (b), double harmonic (c) and triangle (d) trap 
potentials for the same width and depth parameters as those shown in Fig. 
\ref{fig:well_wf}.
Note once again how the solutions to the NLSE satisfy the scaling rule, 
showing the same reduced wavefunction in the reduced units.
For the case of the double harmonic trap and the triangle potential trap [Fig.
(\ref{fig:wfothers}c) and (\ref{fig:wfothers}d)] the wavefunction smooths out 
for small $g_{1,max}R_0$ values in the regions where the potential is 
non-differentiable in contrast to the strong interaction region where the 
wavefunction shows the same shape as the trapping potential.
Thus, we have confirmed for a given defect that two different shapes
with the same $R_0\sqrt{V_0}$ will have the same $g_{1,max}R_0$ and the
same reduced wavefunction.

\begin{figure}[!t]
\includegraphics[width=240pt]{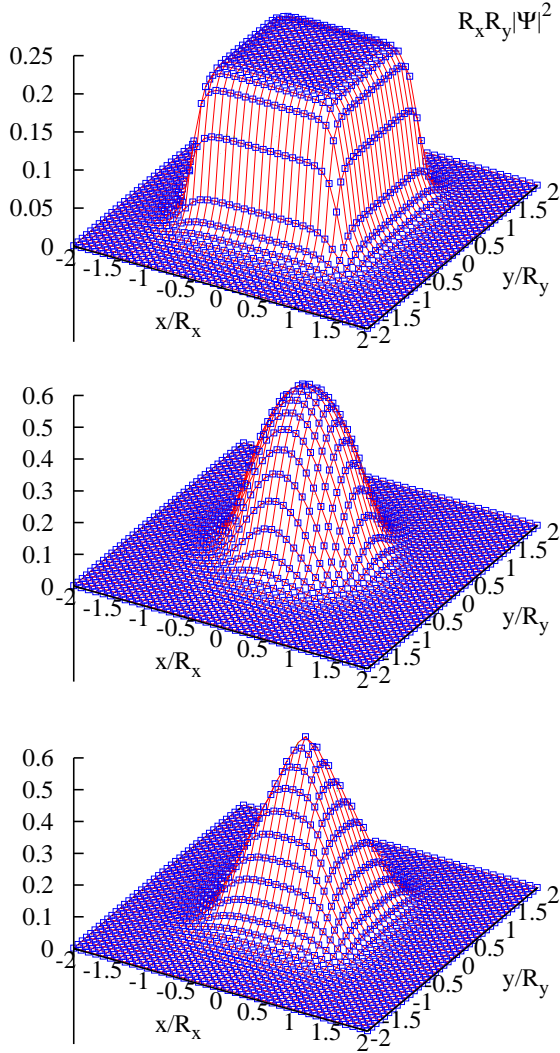}
\caption{\label{fig:2d-wf}
(Color on-line). Numerical scaled density profile solutions,
$|\Phi(\bar{x},\bar{y})|^2$ for the 2-D NLSE as a function
of $\bar{x}$ and $\bar{y}$ for the square, parabolic, and pyramid trap
defects for $V_0=50.0$ and $R_x=R_y=R_0=1.0$ o.u. for the solid lines 
(red lines).
In the same figure we show the results for $V_0=2.0$ and
$R_x=R_y=R_0=5.0$ o.u. represented by the squared symbols (blue), 
thus confirming the scaling density profile rule for the 2-D case.}
\end{figure} 

\subsection{The 2-D trap cases}

For the two-dimensional cases we show in Fig. \ref{fig:2d-wf}
the probability density for the three potential cases (square, harmonic, 
and pyramid well potentials).
In these cases we show the results for $R_x=R_y=R_0=1.0$ o.u. and
$V_0=50.0$ o.u. for the results in red lines (solid lines).
Interestingly, the wavefunction takes a shape similar to the potential the
particles are being held, in the same way as the 1-D case.
In the same figure, we show the results for $R_x=R_y=R_0=5.0$ o.u. and 
$V_0=2.0$ o.u. for the results shown by the symbols (blue squares).
Both cases have $R_0\sqrt{V_0}=7.071$, therefore both of them have the
same scaled wavefunction, $\Phi=\sqrt{R_xR_y}\Psi=R_0\Psi$, confirming our
scaling rule, {\em i.e.} the same reduced density profile.

%

\begin{figure}[!t]
\includegraphics[width=240pt]{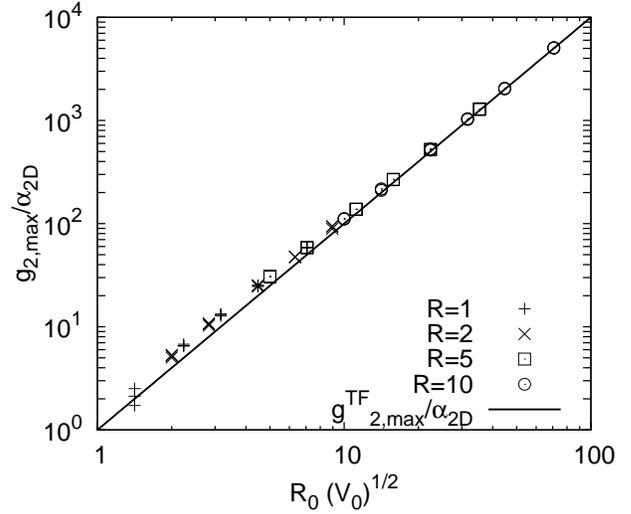}
\caption{\label{fig:g-2d-all}
Scaling law for the 2-D defects, $g_{2,max}/\alpha_{2D}$ of the nonlinear 
coupling constant as a function of $R_0\sqrt{V_0}$ for the square, parabolic, 
and pyramid trapping potential for the 2-D NLSE.
The solid line is the Thomas-Fermi result.
Symbols are our numerical results at the threshold of delocalization.
The upper/middle/lower symbols are, respectively, for square, parabolic
and pyramid shaped defects.}
\end{figure}

In Fig. \ref{fig:g-2d-all}, we show the ratio of the nonlinear coupling
term $g_{2,max}$ to the shape form factor $\alpha_{2D}$ for the 2-D results
as a function of $R_0\sqrt{V_0}$.
The solid lines are the Thomas-Fermi results and the symbols are
the data obtained by our numerical procedure.
Again, for $R_0\sqrt{V_0} > 10$ the Thomas-Fermi results follow closely
the numerical data for $g_{2,max}/\alpha_{2D}$ showing a universal behavior 
for the non-linear coupling term.
Discrepancies start to appear in the results for small values of 
$R_0\sqrt{V_0}$, dependent of the defect shape.  
This is a consequence of the neglect of the kinetic energy term in the TFA, 
and thus the neglect of tunneling for these weakly bound particles.

\subsection{The 3-D trap cases}
\label{sec:3dtrap}

\begin{figure}[!t]
\includegraphics[width=240pt]{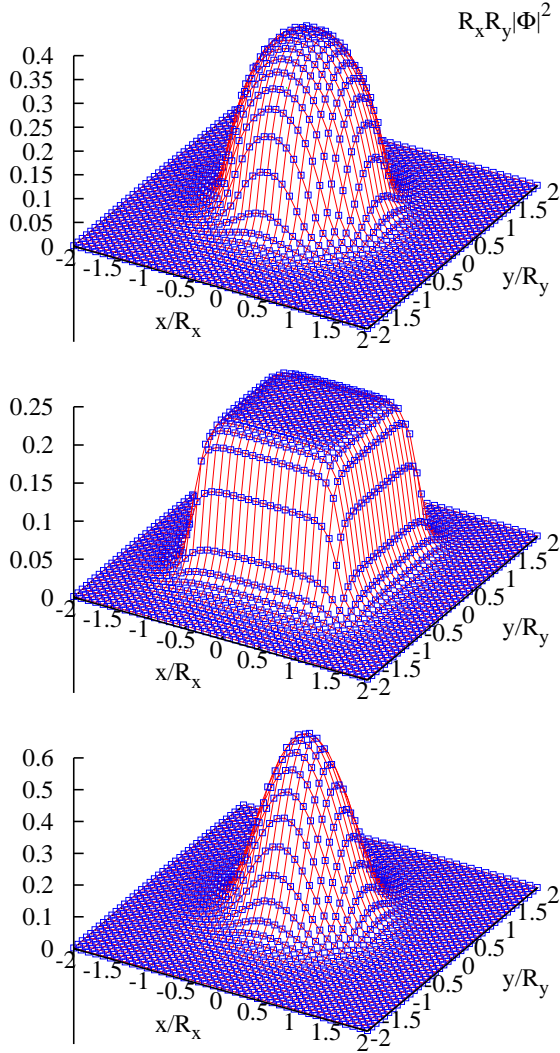}
\caption{\label{fig:3d-wf}
(Color on-line). Numerical projected scaled density profile solutions,
$|\Phi(\bar{x},\bar{y})|^2$ to the 3-D NLSE as a function of $x$ and $y$ 
for the spherical, rectangular cuboid, and parabolic well trap potentials for 
$R_x=R_y=R_z=1.0$ o.u. and $V_0=50.0$ for the solid lines (red lines).
In the same figure we show the results for $R_x=R_y=R_z=2.0$ o.u. and 
$V_0=5.0$ by the squared symbols (blue), thus confirming the scaling
density profile rule for the 3-D case.}
\end{figure}

In order to visualize the density profile of a 3-D wavefunction, we project 
the reduced wavefunction onto the $z$-axis such that
\begin{equation}
|\Phi(\bar{x},\bar{y})|^2=\int |\Phi(\bar{x},\bar{y},\bar{z})|^2 d\bar{z}
\,.
\end{equation}
Thus, for the three-dimensional case, we show in Fig. \ref{fig:3d-wf}
the projected scaled density profiles, $|\Phi(\bar{x},\bar{y})|^2$, for 
the spherical, rectangular cuboid and harmonic well potentials.
For these two cases we show the results for $R_x=R_y=R_0=1.0$ o.u. and
$V_0=50.0$ o.u..  The wavefunction takes a shape similar to the defect
that the particles are being held, as the 1- and 2-D wavefunctions
also tended to do.
In the same figure, we show the results for $R_x=R_y=R_0=5.0$ o.u. and
$V_0=2.0$ o.u..  
Both cases have $R_0\sqrt{V_0}=7.071$, and therefore have the same scaled 
wavefunction $\Phi=\sqrt{R_xR_yR_z}\Psi=R_0^{3/2}\Psi$, as well as the 
same $g_{3,max}/R_0$ confirming that our scaling rule further holds in 3-D.

%

\begin{figure}[!t]
\includegraphics[width=240pt]{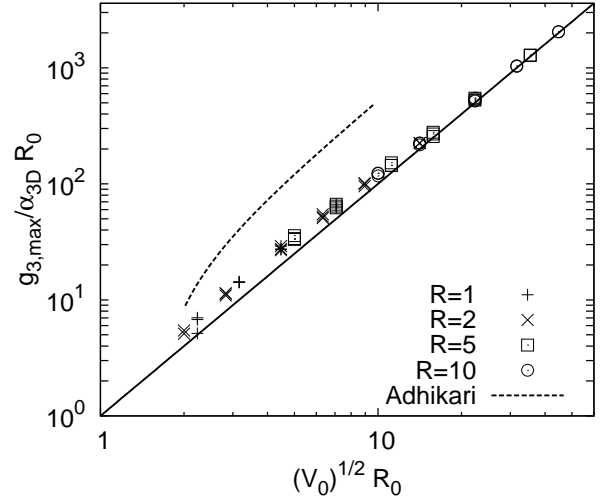}
\caption{\label{fig:g-3d-all}
Scaling law of the 3-D nonlinear coupling constant divided by the potential 
shape factor $\alpha_{3D}$ as a function of $R_0\sqrt{V_0}$ for the
spherical, rectangular cuboid, and parabolic trapping potential for the 3-D 
NLSE.
The solid line is the Thomas-Fermi result.
The symbols are our numerical results near delocalization.
The upper/middle/lower symbols are, respectively, for spherical, 
rectangular cuboid, and parabolic shaped defects.
The dashed line represents the variational results from Adhikari
\cite{Adhikari-SK07-42ejpd279} for the spherical well.}
\end{figure}

In Fig. \ref{fig:g-3d-all}, we show the ratio of the nonlinear coupling
term $g_{3,max}$ to the shape form factor $\alpha_{3D}$ for the three 
dimensional results as a function of $R_0\sqrt{V_0}$.
The solid straight line is the Thomas-Fermi results and together with the 
symbols are the data obtained by our numerical procedure.
Again, for $R_0\sqrt{V_0} > 10$ the Thomas-Fermi results follow closely
the numerical data for $g_{3,max}/\alpha_{3D} R_0$ showing an universal 
behavior for the non-linear coupling constant.
For small values of $R_0\sqrt{V_0}$, discrepancies, dependent of the
potential trap shape, start to appear in the results consequence of the
kinetic energy term. 
In the same figure, we show the variational results for Adhikari
\cite{Adhikari-SK07-42ejpd279}.
As mentioned in the introduction, his results overestimates the TFA or our
numerical results. 

\subsection{Application: A 3-D cube trap}
\label{sec:app}

The application and implications of this work follows directly 
from the scaling law for the non-linear coupling constant and the reduced 
density profile for the 1-, 2-, and 3-D cases considered.
The NLSE in physical units is \cite{Dalfovo-F99-71rmp463}
\begin{eqnarray}
\left\{-\frac{\hbar^2}{2m}\nabla^2+V_{trap}(\mathbf{r})\right.&+& \\
\left. \frac{4\pi \hbar^2 a_s (N-1)}{m}|\Psi_0(\mathbf{r})|^2\right\}
\Psi_0(\mathbf{r}) &=&\varepsilon_0\Psi_0(\mathbf{r}).
\nonumber
\end{eqnarray}
For a trap with characteristic harmonic oscillator lengths $l_x$, $l_y$, and 
$l_z$ for each Cartesian coordinate, such that a transformation to
oscillator units requires $\mathbf{r}=(l_xl_yl_z)^{1/3}
\bar{\mathbf{r}}$, and the NLSE in o.u. becomes
\begin{equation}
\left\{-\frac{1}{2}\bar{\nabla}^2+\bar{V}_{trap}(\bar{
\mathbf{r}})+ g_3
|\Psi(\bar{\mathbf{r}})|^2\right\}\Psi(\bar{\mathbf{r}})=\varepsilon
\Psi(\bar{\mathbf{r}})
\end{equation}
where
\begin{eqnarray}
g_3&=&\frac{4\pi a_s (N-1)}{(l_xl_yl_z)^{1/3}} \\
V&=&\frac{m(l_xl_yl_z)^{2/3}}{\hbar^2}V_3^\prime \\
\varepsilon&=&\frac{m(l_xl_yl_z)^{2/3}}{\hbar^2}\varepsilon_0
\end{eqnarray}

As an example we again choose $^{87}$Rb atoms with $a_s=100\,a_0$ as in
the experiment of Ref. \cite{Anderson-MH95-269science198}.
We consider the trap as a 3-D cube with length scale
$l_x=l_y=l_z=1.222\times 10^{-4}$ cm $=23 000 a_0$
and thus a volume of $8\times l_x^3$.
The TFA then gives us $g_3=0.05464 (N-1)$ and $V=2.68\times 10^8$ $V_{trap}$(o.u/$K$),
where the potential trap depth is given in units of absolute temperature (Kelvin).
Thus, for $R_0$ covering the range from $1$ to $10$ o.u., as used in
Sec. \ref{sec:3dtrap} with physical values that correspond to a BEC with 
extension between $1.22$ and $12.2$ $\mu$m.  Similarly, for $V=V_0$ covering the
range from $1$ to $50$ o.u. this is equivalent to traps depths of $3$ to $150$ nK,
which are values well within the range of the experiment.
The values reported in Fig \ref{fig:g-3d-all} thus correspond to a range 
from $200$ to $2 \times 10^6$ atoms, which are typically found in experiments.

\section{Conclusions}
\label{sec:conclusions}

In conclusion, by means of the Thomas-Fermi approximation at the 
delocalization threshold, $\varepsilon_i=0$, one can easily determine
an approximation to the maximum non-linear interaction strength
$g_{i,max}$ of the Gross-Pitaevskii equation and therefore, the
maximum number of atoms, $N_{max}$, that can be trapped
by $i=$1-, 2-, and 3-D potentials that contain a defect.

The scaling laws of Eqs.~(\ref{eqn:gTFmaxav}), (\ref{eq:g-2d-tf}), and 
(\ref{eq:g-3d-tf}) are dependent on both the depth and length of the 
potential, and remain valid over a surprisingly wide-range of parameters.  
The TFA relies on a wavefunction that is constrained by the
shape of the potential, never diffusing into the classically forbidden
region.  
At the $g_{i,max}$ point the TF wavefunction becomes unbound, and that this 
mimics the actual solution to the NLSE which diffuses towards
infinity as $g_i$ approaches $g_{i,max}$ is remarkable.
It would, however, be worthwhile to extend this work through a
more sophisticated variational treatment 
\cite{Carretero-Gonzalez-R08-21nljR139}.

The existence of these scaling laws is useful because it captures in a
simple expression the number of atoms of a BEC that can be trapped by a
potential defect on waveguide or in free-space.
It would be interesting to further examine the dynamic
scattering/trapping of a BEC as it propagates through such defects
\cite{Ernst-T10-81pra033614,Gattobigio-GL10-12njp085013}.
Essentially, the non-linear term enables a continuum of non-linear bound 
states, as opposed to the quantized single-particle eigenstates.  
For example, if $N_{max} = 200$ atoms, then any number lower than that will 
also be able to be trapped.  
How easy is it for a BEC with $N>N_{max}$ to fill up the defect as it goes 
past remains to be demonstrated, and our initial calculations within the 
NLSE show very little, if any, filling up of the defect.
It will also be interesting to consider that the presence of any
localized atoms will tend to smooth out the defect potential that a
consequent BEC interacting with the defect will experience.
This may have an impact, {\em e.g.} on Anderson-type localization.

\section*{Acknowledgments}

This work has been supported by grants PAPIIT-UNAM IN-101-611 and 
CONACyT-SNI 89607.
RCT acknowledges support from the computer center at ICF-UNAM and
from Reyes Garc{\'\i}a.
MWJB is supported by an ARC Future Fellowship (FT100100905),
and thanks Martin Kandes and Prof. Ricardo Carretero-Gonz{\'a}lez for
helpful correspondence.
BDE acknowledges support from US National Science Foundation.
The germinal research of this project was supported by the
US Department of the Navy, Office of Naval Research.


%

\end{document}